\documentstyle[aps,amssymb,amsmath,pra,epsfig,twocolumn]{revtex} 

\newcommand{\ncd}{\newcommand}
\ncd{\cs}{$|\Phi\rangle_{\cal{C}}\;$}
\ncd{\ncs}{|\Phi\rangle_{\cal{C}}}

\begin{document}
\draft

\title{Persistent entanglement in arrays of interacting particles}

\author{Hans J. Briegel and Robert Raussendorf}

\address{Sektion Physik, Ludwig-Maximilians-Universit\"at, Theresienstr.\ 37,
D-80333 M\"unchen, Germany}

\date{\today}

\maketitle

\begin{abstract}
We study the entanglement properties of a class of $N$ qubit quantum states 
that are generated in arrays of qubits with an Ising-type interaction. 
These states contain a large amount of entanglement as given by their Schmidt 
measure. They have also a high {\em persistency of entanglement} which means 
that $\sim N/2$ qubits have to be measured to disentangle the state.
These states can be regarded as an entanglement resource since one can generate
a family of  other multi-particle entangled states such as the 
generalized GHZ states of $<N/2$ qubits by simple measurements and 
classical communication (LOCC).

\end{abstract}

\medskip


\narrowtext

The notion of entanglement has many facets.  A modern perspective is to regard it as a 
resource for certain communicational and computational tasks \cite{springer-book}. 
Related to this viewpoint is the problem of identifying equivalence classes
of entangled states, and to find relations between these classes.
While for pure states of bi-partite systems there is a single ``unit'' of entanglement -- 
the entanglement contained in a Bell state \cite{bennett95} -- it has recently become clear
that for systems shared by three and more parties there are several inequivalent
classes of entangled states \cite{duer00a,zhang,linden99,duer00b}. Progress in the understanding 
of  multi-particle entanglement has been triggered \cite{duer00a,zhang} by giving 
explicit examples of states that did not fit into existing classification schemes. 
Generally speaking, a sufficiently rich phenomenology of entangled states is needed.
It helps us to refine entanglement classification schemes and, arguably, to motivate 
them in the first place. 

In this paper, we introduce a class of $N$-qubit entangled states which is different
from both the GHZ class and the recently introduced W class of $N$-qubit states \cite{duer00a}. 
We also give an operational characterization of these classes in terms of local measurements. 
In one respect, the states we are going to describe resemble the so-called 
maximally entangled Greenberger-Horne-Zeilinger (GHZ) states \cite{GHZ90} of 
$N$ qubits, while 
in some other respect they are much more entangled than the GHZ states.
To characterize these states, we introduce the notions of 
{\em maximal connectedness} and  {\em persistency of entanglement} of an 
entangled state. 
The first notion emphasizes the possibility that, in an $N$-particle state, 
even when the reduced density matrix of a subset of particles is fully
separable \cite{duer00b}, 
it may still be possible to project that subset of particles into a highly 
entangled state by performing local 
measurements on the {\em other} particles [supplemented with classical
communication and local operations (LOCC), if the parties are remotely
separated]. The second notion relates the amount of entanglement in a
multi-particle  
system to the operational effort it takes (in terms of local operations) to 
destroy all entanglement in the system.
The states we describe occur, for example, in the quantum Ising model
of spin chains  
and, more generally, spin lattices. We will introduce the states in
the context  
of this specific model; their entanglement properties, however, are
discussed in  
general terms by assuming, as usual, that the qubits are distributed between 
remote parties which can only act through LOCC.

Consider an ensemble of qubits that are located on a $d$-dimensional lattice 
($d=1,2,3$) at sites $a\in  
{\mathbb Z}^d$ and interact via some 
short-range interaction described by the Hamiltonian
\begin{equation}
H_{\text{int}}= \hbar g(t) \sum_{a,a'} f(a-a') \frac{1+\sigma_z^{(a)}}{2}
\frac{1-\sigma_z^{(a')}}{2}
\label{collision}
\end{equation}
Concerning the entanglement properties of the states we are going to investigate,
this interaction Hamiltonian is equivalent to the quantum Ising model with 
$H_{\text{int}}'= 
- \sum_{a,a'}\frac{1}{4}\hbar g(t) f(a-a') \sigma_z^{(a)}\sigma_z^{(a')}$, where 
the indices $a,a'$ run over all occupied lattices sites. 
The coupling strength is written as a product $gf$, where $f(a-a')$ specifies the 
interaction range and the $g(t)$ allows for a possible overall time dependence. 
In this letter, we confine ourselves to next-neighbor interactions. A more general 
situation will be reported in \cite{raussen}.  
In the language of quantum information, the interaction (\ref{collision}) 
realizes simultaneous conditional phase gates between qubits at neighboring sites 
$a$ and $a'$. For an experimental realization see the discussion at the end of the 
paper.

Consider first the one-dimensional example of a chain of $N$ qubits (``spin chain'') 
\cite{not-wootters} with next-neigbor interaction $f(a-a')=\delta_{a+1,a'}$. 
Initially, all qubits are prepared in the state
$(|0\rangle_a + |1\rangle_a)/\sqrt{2}$, where $|0\rangle_a\equiv |0\rangle_{z,a}$ and
$|1\rangle_a=|1\rangle_{z,a}$ are eigenstates of $(1-\sigma_z^{(a)})/2$ with eigenvalues 
$0$ and $1$, respectively. (This is the most interesting situation; if they are 
prepared in states $|0\rangle_a$ or $|1\rangle_a$, no entanglement will build up.) 
The unitary transformation generated by (\ref{collision}) is 
$U(\varphi)=\exp(-{\rm i} \varphi \sum_a \frac{1+\sigma_z^{(a)}}{2}
\frac{1-\sigma_z^{(a+1)}}{2})$ with $\varphi=\int\text{d}t g(t)$. 
For $g(t)=g=$const, $U(\varphi)=
U(gt)$ is periodic in time and generates ``entanglement oscillations"
of the chain. For the specific values $\varphi=0,2\pi,4\pi,\dots$ the chain is 
disentangled, while for all other values of $\varphi$, it is entangled. 
For the values $\varphi = \pi,3\pi,5\pi,\dots$ the chain is in some sense 
maximally entangled and we will concentrate on this situation in the following. 
The state can then be written in the form
\begin{equation}
|\phi_N\rangle = \frac{1}{2^{N/2}} \bigotimes_{a=1}^{N} 
\left(|0\rangle_a \sigma_z^{(a+1)}+|1\rangle_a \right)
\label{spin-chain}
\end{equation}
with the convention $\sigma_z^{(N+1)}\equiv 1$. The compact notation 
employed in (\ref{spin-chain}) is easily understood by multiplying out 
the right hand side. For $N=2$, one obtains $|\phi_2\rangle =\frac{1}{2}
(|0\rangle_1 \sigma_z^{(2)}+|1\rangle_1)
(|0\rangle_2 +|1\rangle_2)
=\frac{1}{2}\left( |0\rangle_1(|0\rangle_2 - |1\rangle_2)
+ |1\rangle_1(|0\rangle_2 + |1\rangle_2) \right)$ which is a maximally  
entangled state. We may write it, up to a local unitary transformation
on qubit 2 in the standard form
\begin{equation}
|\phi_2\rangle =_{\scriptsize l.u.} \frac{1}{\sqrt{2}}(|0\rangle_1|0\rangle_2
+|1\rangle_1|1\rangle_2)\,,
\label{2qubits}
\end{equation}
where ``l.u." indicates that the equality holds up to a local unitary 
transformation on one or more of the qubits \cite{linden98}. 
Similarly, one obtains for $N=3,4$
\begin{eqnarray}
|\phi_3\rangle &=_{\scriptsize l.u.} & 
\frac{1}{\sqrt{2}}\left(|0\rangle_1|0\rangle_2|0\rangle_3
+ |1\rangle_1|1\rangle_2|1\rangle_3\right)\,, \nonumber \\ 
|\phi_4\rangle &=_{\scriptsize l.u.} & 
\frac{1}{2}\left(|0\rangle_1|0\rangle_2|0\rangle_3|0\rangle_4
+ |0\rangle_1|0\rangle_2|1\rangle_3|1\rangle_4 \right. \nonumber \\
&& \left.
+ |1\rangle_1|1\rangle_2|0\rangle_3|0\rangle_4
- |1\rangle_1|1\rangle_2|1\rangle_3|1\rangle_4\right)\,.
\label{3and4qubits}
\end{eqnarray}
While $|\phi_3\rangle$ corresponds to a GHZ state of three qubits
\cite{GHZ90}, $|\phi_4\rangle$ is {\em not equivalent} to a 4-qubit  
GHZ state \cite{zhang}. More generally, the states $|\phi_N\rangle$ and the 
$N$-qubit GHZ state $|{\sc GHZ}_N\rangle \equiv 2^{-1/2}(|0\rangle_1\dots
|0\rangle_N) + |1\rangle_1\dots |1\rangle_N)$ are not equivalent for $N>3$ i.e.\ 
cannot be transformed into each other by LOCC (local transformations and classical 
communication) as we shall see below.

How can we compare the entanglement properties of $|\phi_4\rangle$ and 
$|\sc{GHZ}_4\rangle$ in operational terms? Imagine that the qubits are distributed 
between four remote parties, which may perform, as usual, local operations 
and classical communication. We observe: a) The states 
{\em share} the property that any two of the four qubits can be projected 
into a Bell state by measuring the other two qubits in an appropriate basis. 
In other words, the parties may use either of the states  
$|\phi_4\rangle$ or $|\sc{GHZ_4}\rangle$ to {\em teleport} \cite{bennett93} 
a qubit between 
any of the four parties. b) The states are {\em different} in that it
is harder  
to {\em destroy} the entanglement of state $|\phi_4\rangle$ than that of 
$|\sc{GHZ_4}\rangle$ by local operations. In fact, {\em it is impossible to 
destroy all entanglement of $|\phi_4\rangle$ by a single local operation,
such as a von Neumann measurement or complete depolarization of a qubit.}
For the state $|\sc{GHZ_4}\rangle$, in contrast, a single local 
measurement suffices to bring it into a product state \cite{gisin98}.

These observations motivate us to introduce the following definitions.
A local measurement in the following means a von Neumann measurement 
on a single qubit.  
\newline {\bf Definition 1:} (Max. connectedness) The quantum mechanical 
state of a set ${\cal C}=\{1, 2, \dots, n\}$ of $n$ qubits is {\em maximally 
connected} if any two qubits $j\ne k\in {\cal C}$ can be projected, with certainty,
into a pure Bell state by local measurements on a subset of the other qubits.    
\newline
Note that the state obtained may depend on the outcome of the measurements.
\newline {\bf Definition 2:} (Persistency) The {\em persistency of entanglement} 
$P_e$ of an entangled state of $n$ qubits is the minimum number of local measurements 
such that, for all measurement outcomes, the state is completely {\em disentangled}.  
\newline
Since we are only concerned with pure states, a disentangled state 
means a product state of all $n$ qubits \cite{terno99}.
Obviously, for all $n$-qubit states $0\le P_e\le n-1$. 

Definitions 1 and 2 can be straightforwardly generalized 
to arbitrary $n$-partite pure states.
Note that the definitions 1 and 2 are invariant under the group of
local unitary  
transformations on any of the qubits \cite{linden98}.

In the sense of these definitions, both states $|\phi_4\rangle$ and 
$|\sc{GHZ_4}\rangle$ are maximally connected, while their persistency is $P_e=2$ 
and $P_e=1$, respectively. More generally, for the state $|\phi_N\rangle$ we 
show that (i) it is maximally connected and
(ii) its persistency is  $P_e(|\phi_N\rangle)=  \lfloor N/2 \rfloor$. 
Property (ii) quantifies the operational effort that is 
needed to destroy all entanglement in the qubit chain. 
We also note that, (iii), the persistency of the states $|\phi_N\rangle$ is equal 
to their Schmidt measure \cite{eisert00}: If one 
expands $|\phi_N\rangle$ into a product basis of the $N$ qubits, the 
minimum number of terms in such a generalized Schmidt representation 
\cite{duer00a,eisert00} grows {\em exponentially} and requires $2^{\lfloor N/2 \rfloor}$ 
product terms. In that sense, the state $|\phi_N\rangle$ of the qubit chain is indeed 
much more entangled than most of the known $N$ qubit states.

We now prove property (i). The cases $N=2,3$ are trivial as the state
is a Bell  
or a GHZ state, respectively. For $N >3$, the proof goes as follows. Let us denote
by $|0\rangle_{xj}\equiv (|0\rangle_j+|1\rangle_j)/\sqrt{2}$, 
$|1\rangle_{xj}\equiv (|0\rangle_j-|1\rangle_j)/\sqrt{2}$ the eigenstates of 
$\sigma_x^{(j)}$. We first show that the qubits at the ends
of the string, i.e., qubits $1$ and $N$ can be brought into a Bell state by measuring 
the qubits $2, \dots, N-1$. For easier book keeping, we use the notation 
$|\phi_N\rangle \equiv |\{1,2,3,\dots,N\}\rangle_{\text{chain}}$. Then the state can be 
expanded in the form $|\{1,2,3,\dots,N\}\rangle_{\text{chain}}= (|0\rangle_1\sigma_{z}^{(2)} 
+|1\rangle_1)(|0\rangle_2\sigma_{z}^{(3)} +|1\rangle_2)|\{3,4,\dots,N\}\rangle_{\text{chain}}$
where we suppress normalization factors. Measuring the operator $\sigma_x^{(2)}$ 
of qubit 2, we obtain for the remaining (unmeasured) qubits 
$1,3, 4, \dots , N$ the state $ _{x2}\langle \epsilon_2|\{1,2,3,\dots,N\}
\rangle_{\text{chain}} = \{(1-i\sigma_y^{(1)})/\sqrt{2},(\sigma_x^{(1)}+\sigma_z^{(1)})/\sqrt{2}\}$
$|\{1,3,4,\dots,N\}\rangle_{\text{chain}}$ for the outcome 
$\epsilon_2 = \{0,1\}$, correspondingly.
This state is, up to the local unitary transformations specified in the parenthesis, 
identical to an entangled chain of length $N-1$, and gives us a recursion formula.
We can repeat this procedure and measure qubit 3, and so on.
We obtain $\bigotimes_j$$(_{xj}\langle\epsilon_j|)|\{1,2,3,\dots,N\}
\rangle_{\text{chain}}=U_1 |\{1,N\}\rangle_{\text{chain}}$ with  $U_1 \in 
\{1,\sigma_{x}^{(1)},\sigma_{y}^{(1)},\sigma_{z}^{(1)}\}$ for $N$ even
and $U_1 \in \{(\sigma_{x}^{(1)}\pm \sigma_{z}^{(1)})/\sqrt{2},
(1\pm i\sigma_{y}^{(1)})/\sqrt{2}\}$ for $N$ odd, up to a phase factor. This is a Bell 
state.
To bring any other qubits $j,k$ (w.l.o.g. $j<k$ ) from the chain $\{1,2,\dots, N\}$ into a
 Bell 
state, we first measure the ``outer" qubits $1,2,\dots j-1$ and  $k+1, k+2,\dots, N$ 
in the $\sigma_z$ basis, which projects the qubits of the remaining chain $j,j+1,\dots, 
k$
into the state $U_j\otimes U_k |\{j,j+1,\dots,k-1,k\}\rangle_{\cal C}$ with $U_j \in \{1, 
\sigma_z^{(j)}\}$, $U_k \in \{1, \sigma_z^{(k)}\}$.  A subsequent measurement of the 
``inner" 
qubits  $j+1, \dots k-1$ will then project qubits $j,k$ into a Bell state, as shown 
previously.

To prove property (iii), we use the expansion 
\mbox{$|\{1,2,3,\dots,N,N+1,N+2\} \rangle_{\text{chain}}=$} 
$|\{1,2,3,\dots,N\}\rangle_{\text{chain}}(|0\rangle_{N+1}-|1\rangle_{N+1}\sigma_z^{(N)})$
$(|0\rangle_{N+2}-|1\rangle_{N+2}\sigma_z^{(N+1)})$ which can be written in the form
$|\phi_{N+2}\rangle= |\phi_{N}\rangle |0\rangle_{x,N+1}|1\rangle_{z,N+2}
- (\sigma_z^{(N)} |\phi_{N}\rangle) |1\rangle_{x,N+1}|0\rangle_{z,N+2}$.
Denote the minimum number of product terms in an expansion of $|\phi_{N}\rangle$
by $r$. As this number is invariant under local unitary transformations 
\cite{duer00a,eisert00}, it is the same for the state $\sigma_z^{(N)} 
|\phi_{N}\rangle$. 
No term in an expansion of $|\phi_{N}\rangle |0\rangle_{x,N+1}|1\rangle_{z,N+2}$ 
can be combined with any term in an expansion of  $(\sigma_z^{(N)} |\phi_{N}\rangle) 
|1\rangle_{x,N+1}|0\rangle_{z,N+2}$ 
into a single product term, since any nontrivial linear combination of 
$|0\rangle_{x,N+1}|1\rangle_{z,N+2}$ with $|1\rangle_{x,N+1}|0\rangle_{z,N+2}$
gives a non-product state w.r.t.\ qubit $N+1$ and $N+2$. 
The minimum number of product terms for an expansion of $|\phi_{N+2}\rangle$ is thus
equal to $2r$. Since for $N=2,3$ we have $r=2$ [see (\ref{2qubits}) and 
(\ref{3and4qubits})], it follows by induction that $r=2^{\lfloor N/2\rfloor}$. 
In other words, the Schmidt measure $P_{\text{S}}(|\phi_N\rangle)$
\cite{eisert00} of $|\phi_N\rangle$ is equal to $\log_2(r)=\lfloor N/2\rfloor$.

We now prove property (ii). An explicit strategy to disentangle state (\ref{spin-chain})
is to measure $\sigma_z^{(j)}$ of all even numbered qubits, $j=2,4,6,\dots $, which can easily
be verified. The total number of these measurements is $\lfloor N/2\rfloor$, which gives an upper 
bound to the persistency, i.e. $P_e(|\phi_N\rangle) \le \lfloor N/2\rfloor$. 
On the other hand, the Schmidt measure gives a lower bound to the persistency.
This can be seen as follows. Since $|\phi_N\rangle$ can be disentangled
by $P_e$ measurements, there exists an expansion of the form
$|\phi_N\rangle = \sum_{j_1,\dots,j_{P_e}=0}^{1}|\mu_1^{(j_1)}\rangle_{a_1}|\mu_2^{(j_2)}\rangle_{a_2} 
\dots |\mu_{P_e}^{(j_{P_e})}\rangle_{a_{P_e}}|{\rm prod}^{(j_1,\dots,j_{P_e})}\rangle$ where 
$a_1,\dots, a_{P_e}$ are the measured atoms, $|\mu_1^{(j_1)}\rangle_{a_1}, \dots,
|\mu_{P_e}^{(j_{P_e})}\rangle_{a_{P_e}}$ the resulting 1-qubit states 
for the measurement outcomes $j_1,j_2,\dots, j_{P_e}$, 
and $|{\rm prod}^{(j_1,\dots,j_{P_e})}\rangle$ 
some (unnormalized) product states of the remaining qubits. This expansion contains at most $2^{P_e}$
product terms, and therefore $P_{\text{S}}\le \log_2(2^{P_e}) = P_e$.
Together with (iii) we obtain $P_{\text{S}}(|\phi_N\rangle) =  \lfloor N/2\rfloor 
\le P_e(|\phi_N\rangle) \le 
\lfloor N/2\rfloor$ which proves property (ii).  

Results (ii) and (iii) show that the persistency of entanglement of the state $|\phi_N\rangle$ 
(\ref{spin-chain}) coincides with its Schmidt measure. This result also holds for the state 
$|{\sc GHZ}_N\rangle$. The meaning of these two concepts is, however, not the same. 
To illustrate this point, consider the so-called W state discussed in Ref.~\cite{duer00a},
$|W_N\rangle =  N^{-1/2}(|1\rangle_1|0\rangle_2 \dots
|0\rangle_N + |0\rangle_1|1\rangle_2 \dots
|0\rangle_N  + \dots + |0\rangle_1|0\rangle_2 \dots
|1\rangle_N)$. The Schmidt measure of this state is equal to $\log_2(N)$ which means that the amount 
of entanglement contained in $|W_N\rangle$ is smaller, in fact exponentially smaller, than in the state 
$|\phi_N\rangle$. The persistency of $|W_N\rangle$, on the other hand, is given by $N-1$ 
\cite{gracias_simon_hardy} which means that the entanglement of $|W_N\rangle$ is harder to destroy
by local measurements than that of $|\phi_N\rangle$. This observation agrees with the findings of 
Ref.~\cite{duer00a}, who showed that any state obtained from   $|W_N\rangle$ by tracing over
$N-2$ qubits is inseparable. Note however that, different from $|\phi_N\rangle$ and $|{\sc GHZ}_N\rangle$,
the state  $|W_N\rangle$ is not maximally connected.

\begin{figure}[tbp]
\hspace*{-0.25cm}
\epsfig{file=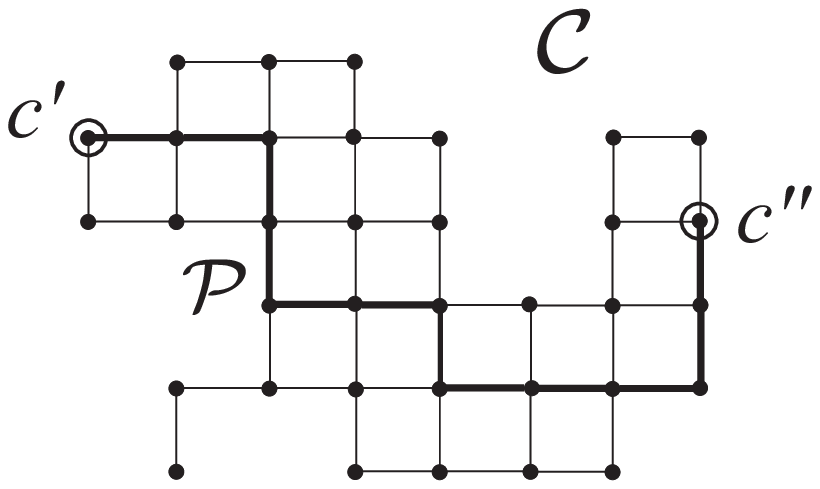,width=4.5cm}\hspace{0.65cm}
\epsfig{file=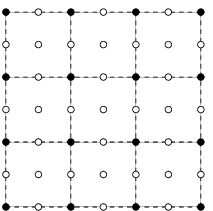,width=2.5cm}
\caption{(a) Entangled cluster $\cal C$ of two-state particles. Any two qubits $c', c''$ of the 
cluster  may be projected into a Bell state by measurements on other qubits of the cluster.
(b) Generation of a 16 qubit generalized  GHZ state from state (\ref{2D-cluster3}) of a 2D block 
    of 49 qubits. The GHZ state is simply obtained by measuring the circled qubits in the appropriate 
    basis [$\sigma_x$ on the lines; $\sigma_z$ between the lines], and by subsequent 1-qubit rotations
    on the remaining qubits.}
\label{FIGcluster}
\end{figure}

In the remainder of the paper we will generalize some of the results to dimensions
$d=2$ and $d=3$, i.e.\ to qubits arranged on a lattice. 
These cases are different from the case $d=1$ since there is no 
natural ordering of the qubits. Therefore, the concept of a ``chain'' of qubits 
does not apply anymore. The natural generalization to higher
dimensions is a  ``cluster'' $\cal C$ of qubits as in Fig.~\ref{FIGcluster}a.
The precise definition of a cluster is the following: Let each lattice site be specified
by a $d$-tuple  of (positive or negative) integers $a\in {\mathbb Z}^{d}$. 
Each site $a$ has $2d$ neighbouring sites. If occupied, these are the sites whose qubit 
interacts with the qubit at $a$.  The set  ${\cal A}\subset {\mathbb Z}^d$ 
specifies all sites that are occupied by a qubit. Two sites $a,a' \in {\cal A}$ 
are {\em connected} (in a topological sense) if there exists a sequence of 
neighboring sites that are all occupied, that is 
$\{a^{(n)}\}_{n=1}^{N}\subset {\cal A}$ with  
$a^{(1)}=a$ and $a^{(n)}=a'$.  A {\em cluster} ${\cal C}\subset {\cal A}$ is a 
subset of ${\cal A}$ with the properties  that first, any two sites 
$c,c'\in {\cal C}$ are connected  and second, any sites $c \in {\cal C}$ and 
$a \in {\cal A}\backslash {\cal C}$ are not connected. 

The quantum mechanical state of a cluster that is generated under the Hamiltonian 
(\ref{collision}) for $\varphi=\pi$ is 
\begin{equation}
    |\Phi\rangle_{\cal C} = \bigotimes_{c\in {\cal C}} \left( 
    |0\rangle_{c}\bigotimes_{\gamma \in \Gamma}\sigma_z^{(c+\gamma)} 
    +|1\rangle_{c}\right) 
   \label{2D-cluster3}
\end{equation}
with the choice $\Gamma = \{(1,0),(0,1)\}$ for $d=2$ and 
$\Gamma = \{(1,0,0),(0,1,0),(0,0,1)\}$ for $d=3$, using the
convention that $\sigma_z^{(c+\gamma)}\equiv1$ when 
$c+\gamma \notin {\cal C}$ (the qubit cannot be entangled with an
empty site). The special case of the 1D chain (\ref{spin-chain}) is obtained 
from (\ref{2D-cluster3}) for the choice  $\Gamma = \{1\}$.  

The cluster states (\ref{2D-cluster3}) satisfy the following set of eigenvalue equations:
\begin{equation}
    K_a |\Phi\rangle_{\cal C} = \kappa |\Phi\rangle_{\cal C} 
    \label{correlation-operators}
\end{equation}  
for the family of operators $K_a=\sigma_x^{(a)}$ $\bigotimes_{\gamma \in \Gamma \cup -\Gamma}$
$\sigma_z^{(a+\gamma)}$,   $a\in {\cal C}$, where $\Gamma \cup -\Gamma $ specifies the 
sites of all qubits that interact with $a$, and  $\sigma_z^{(a+\gamma)}\equiv1$ when  
$a+\gamma \notin {\cal C}$. The eigenvalue $\kappa = \pm 1$ is determined by the specific 
occupation pattern of the neighboring sites. For $a+\{\Gamma \cup -\Gamma\}\subset {\cal C}$, 
for example, $\kappa=(-1)^d$. The operators  $\{K_a | a\in {\cal C}\}$ form a complete 
set of commuting observables of which the cluster states \cs are are eigenstates.

Equations (\ref{correlation-operators}) can be used to generalize some of the entanglement 
properties from the 1D case to higher dimensions. Here we just report the results. 
A detailed proof will be given in a longer paper \cite{raussen}.

We find that all cluster states are maximally connected. It is noteworthy that
the property of maximal connectedness of \cs does 
not depend on the precise shape of the cluster, and not even on its topological 
characterization except for being a cluster. Consider a cluster 
$\cal C$ and any two qubits on sites $c', c''\in {\cal C}$ as in Fig~\ref{FIGcluster}a. 
To bring these qubits into a Bell state, we first 
select a {\em one-dimensional path} ${\cal P}\subset {\cal C}$ that connects sites $c'$ and 
$c''$ as in Fig.~\ref{FIGcluster}a. Then we measure all neighboring qubits 
surrounding this path in the $\sigma_z$ 
basis.
\noindent  By this procedure, we project the qubits on path  $\cal P$ into a state that is, 
up to local unitary transformations, identical to the state $|\Phi_N\rangle$ of the linear 
chain. We have thereby reduced the two- and three-dimensional problem to the one-dimensional problem.

Equations (\ref{correlation-operators}) can also be used to  calculate the persistency, as they
imply strict correlations among 1-particle measurements. These correlations can be used to 
minimize the number of measurements required to project \cs into a product state. 
In general, the exact value of the persistency depends on the shape of a cluster. 
For large convex clusters, we can give the asymptotic result 
$P_e/N = 1/2$ where $N\to\infty$ is the number of qubits.

Entanglement is often regarded as a resource and thus the question arises which 
states can be obtained from cluster states by  local operations and classical 
communication (LOCC). A particularly simple class of LOCC is obtained by restricting 
oneself to projective von Neumann measurements on selected qubits. We note without 
proof \cite{raussen} that from a  block $\cal C$
of $L^d$ qubits, one can obtain any state of the form 
$\alpha|00\dots 0\rangle_{\cal C'}
+\beta |11 \dots 1\rangle_{\cal C'}$ of any subset of qubits ${\cal C}'\subset 
{\cal C}\cap 
\{2{\mathbb Z}\}^d$. For $\alpha=\beta=1/\sqrt{2}$, this includes, 
in particular, the family of generalized (multi-particle) GHZ states on this subset. An illustration 
is given in Fig.~\ref{FIGcluster}b.
Even though the thereby obtained states are highly entangled, their Schmidt entanglement 
measure \cite{eisert00} is always smaller than of the original cluster state, and so 
the total amount of entanglement decreases.
  
With the experimental progress in cooling and trapping of neutral atoms, one has identified 
systems such as ``optical lattices'' \cite{brennen99,jaksch99} in which the interaction 
(\ref{collision}) can be implemented by cold atomic collisions \cite{jaksch99} or other techniques. 
These systems allow one, in particular,
to switch on and off the coupling $g(t)$ between all qubits simultaneously by a manipulation of the
parameters of the trapping lasers. The unitary transformation $U(\varphi)$ 
[before eq.(\ref{spin-chain})] with $\varphi=\pi$ can thereby be realized by a single global operation. 
This enables one, in principle, to create a variety of multi-particle entangled states
such as $|{\sc GHZ_M}\rangle$ with $M\gg 3$ by the entanglement operation $U(\varphi)$, 
followed by 1-qubit measurements and subsequent 1-qubit rotations (compare Fig.~\ref{FIGcluster}b).

In conclusion we have introduced a class of highly entangled multi-qubit states.
The cluster states have a large persistency of entanglement which quantifies the operational 
effort needed to disentangle these states. For the chain of qubits in the state 
$|\phi_N\rangle$, we have shown that the value of the persistency agrees with the
Schmidt measure of $|\phi_N\rangle$. In that sense, the state $|\phi_N\rangle$ is
indeed much more entangled than most known $N$ qubit states. 
The cluster states can be regarded as a (scalable) resource for other multi-qubit 
entangled states, such as multiparticle GHZ states. Experimentally, these states could 
be generated and studied in optical lattices or similar systems.

We thank H. Aschauer, B.-G.\ Englert, J.\ Hersch, L.\ Hardy, A.\ Schenzle, and C.\ Simon 
for discussions. One of us (HJB) enjoyed delightful discussions with Jens Eisert that 
emerged from a hiking tour during the Benasque workshop 2000. We are also grateful to 
J. Eisert for comments on the manuscript. This work has been supported 
in part by the Schwerpunktsprogramm QIV of the DFG.



\begin{references}
\bibitem{not-wootters}
The interaction we are considering is different from the spin chain 
studied by Wootters in \cite{wootters}.
\end{references}
\end{document}